
\newskip\oneline \oneline=1em plus.3em minus.3em
\newskip\halfline \halfline=.5em plus .15em minus.15em
\newbox\sect
\newcount\eq
\newbox\lett
\newdimen\short
\def\adv{\global\advance\eq by1}
\def\set#1#2{\setbox#1=\hbox{#2}}
\def\nextlet#1{\global\advance\eq by-1\setbox
                \lett=\hbox{\rlap#1\phantom{a}}}

\newcount\eqncount
\eqncount=0
\def\equn{\global\advance\eqncount by1\eqno{(\the\eqncount)}}
\def\put#1{\global\edef#1{(\the\eqncount)}           }

\def\mbox#1#2{\vcenter{\hrule \hbox{\vrule height#2in
                \kern#1in \vrule} \hrule}}  
\def\sq{\,\raise.5pt\hbox{$\mbox{.09}{.09}$}\,}
\def\sqb{\,\raise.5pt\hbox{$\overline{\mbox{.09}{.09}}$}\,}

\def\sig{\sigma}
\def\nabar{{\overline \nabla}}
\def\bR{\overline R}
\def\R{R^2}
\def\Ric{R_{ab}R^{ab}}
\def\Rie{R_{abcd}R^{abcd}}
\def\sqR{\sq R}
\def\eps{\epsilon}

\def\trp{\rm{Tr'}}

\def\h1{\hat 1}

\def\cgrav{1}
\def\cAntMot{2}
\def\cPoly{3}
\def\cAM{4}
\def\cAMM{5}
\def\cselfd{6}
\def\cDuf{7}
\def\cwz{8}
\def\cZam{9}
\def\cmes{10}
\def\cDDK{11}
\def\cBer{12}

\magnification=1200
\hsize=6.0 truein
\vsize=8.5 truein
\baselineskip 14pt

\nopagenumbers

\rightline{CPTH-C207.1192}
\rightline{November 1992}
\vskip 1.5truecm
\centerline{{\bf DYNAMICS OF THE CONFORMAL FACTOR IN 4D GRAVITY}
\footnote\dag{Invited talk at the Spring Workshop on String Theory,
Trieste, Italy, April 8-10, 1992}}
\vskip 1.5truecm
\centerline{\bf I. Antoniadis}
\vskip .5truecm
\centerline{\it Centre de Physique Th{\'e}orique}
\centerline{\it Ecole Polytechnique, 91128 Palaiseau, France}
\vskip 1.5truecm
\centerline{\bf ABSTRACT}
\vskip .5truecm
We argue that 4D gravity is drastically modified at distances larger
than the horizon scale, due to the large infrared quantum fluctuations
of the conformal part of the metric. The infrared dynamics of the
conformal factor is generated by an effective action, induced by the
trace anomaly of matter in curved space, analogous to the Polyakov
action in two dimensions. The resulting effective scalar theory is
renormalizable, and possesses a non-trivial, infrared stable fixed
point, characterized by an anomalous scaling dimension of the conformal
factor. We argue that this theory describes a large distance scale
invariant phase of 4D gravity and provides a framework for a dynamical
solution of the cosmological constant problem.
\vskip 4truecm

\hfill\break
\vfill\eject

\footline={\hss\tenrm\folio\hss}\pageno=1

The cosmological constant, if not identically zero, is the smallest
fundamental mass scale in nature. In contrast to other fine-tuning
problems, an adjustment of its value once is not sufficient to explain
its smallness at all epochs in the evolution of the universe. Thus,
any attempt at an explanation cannot rely on an exact symmetry of the
fundamental quantum theory of gravitation (such as supersymmetry), if
that symmetry is ultimately broken in the low energy effective
theory. Instead, the fact that the effective cosmological ``constant"
is dynamically dependent on the vacuum state of all quantum fields in
nature implies that not the physics of the Planck scale, but the
low-energy, or {\it infrared} dynamics of gravity is essential to a
resolution of the problem.

This conclusion seems to contradict the naive expectation that
at distances larger than the Planck length gravity is very well
described by the classical Einstein theory. There are indeed two
main reasons that could invalidate this expectation:
\hfill \break
(i) On the one hand, earlier studies of quantum fields in curved
spacetime made in evidence peculiar long distance dynamics for massless
fields in rapidly expanding cosmological backgrounds, such as de	Sitter
space which is the natural maximally symmetric ground state of Einstein
gravity with a positive cosmological term. In particular, the
graviton propagator grows without bound at distances larger than the
horizon length [\cgrav ,\cAntMot], while the spin zero or conformal
part of the propagator provides the dominant contribution [\cAntMot].
This suggests that classical theory is only valid at intermediate
distances, larger than the Planck scale but smaller than the horizon,
and that in the far infrared regime quantum gravitational effects
become important.
\hfill \break
(ii) On the other hand, the existence of trace anomaly implies that the
trace of the classical equations of motion cannot in general be imposed
consistently with general covariance at the quantum level. A clear
example is provided by the situation in two dimensions, where the
quantum trace anomaly changes discontinuously the classical theory
and generates an effective non-local action for 2D gravity [\cPoly].
In the conformal gauge, this action becomes a local kinetic term for
the conformal part of the metric, the so called Liouville mode, whose
dynamics determines the non-trivial critical behavior of 2D gravity.

We are led to study the infrared dynamics of 4D gravity, at distances
larger than the horizon. This region is expected to be insensitive to
the ultraviolet structure of the theory. To find the relevant effective
action, one could therefore put an ultraviolet cutoff at the horizon
length and integrate all high frequency modes up to that scale. At such
large distances with several causally disconnected regions, all matter
interactions can be neglected, while massive matter being red-shifted
away very rapidly is decoupled. The remaining massless fields could
then be treated as free and classically conformally invariant. The
metric can be parametrized in conformal coordinates
$$
g_{ab}(x) = e^{2 \sig(x)} \bar g_{ab}(x), \equn\put\confdef
$$
where the conformal factor $e^{2 \sig(x)}$ is factorized and
$\bar g_{ab}(x)$ contains the transverse spin-2 excitations. At a first
approximation, one can treat $\bar g_{ab}(x)$ as a fixed fiducial
metric and study the infrared dynamics of the conformal factor.
Although in the classical theory the conformal part of the metric is
completely constrained, the situation changes dramatically at the
quantum level due to the trace anomaly. Remarkably, the
$\sig$-dependence of the induced non-local action turns out to be
local, as in two dimensions, and an exact analysis of the infrared
critical behavior of the theory can be performed [\cAM].

Although in two dimensions the freezing of $\bar g_{ab}$ corresponds
to a choice of gauge, in four dimensions it represents a severe
truncation of the full theory. However, there are good reasons to
believe that this could be still a good approximation in order to study
the  long distance dynamics: In fact, the conformal factor gives rise to
the most singular infrared behavior in perturbation theory around
de Sitter space. Its classical equations of motion, which fix the
curvature  scalar to be constant, are modified by the quantum trace
anomaly. Finally, it plays a dominant role in cosmology and the
truncated theory is already much richer than the ``minisuperspace"
truncation to a finite number of degrees of freedom. After
all, the transverse spin-2 excitations could be treated as the
remaining massless matter fields, which just amounts to provide their
own contribution to the anomaly coefficients [\cAMM].

Appart from the cosmological constant problem, the dynamics
of the conformal factor may have other interesting applications, as
in inflationary cosmology, large-scale structure, and the dark matter
problem. Moreover, it is an interesting non-trivial field theory in
its own, which generalizes the Liouville theory in four dimensions and
describes an exact treatment of 4D self-dual gravity [\cselfd].
\vskip 1.0cm
\centerline{\bf The effective theory}
\vskip 0.5cm

In order to determine the effective action, which describes the
large-distance dynamics of the conformal factor, we proceed in
analogy with the two-dimensional case which we review below. The
classical action of 2D gravity contains no derivatives of the metric
since the Einstein-Hilbert term is a total derivative:
$$
\eqalign{S^{(2)}_{cl} &= \int \sqrt{-g}\ (\gamma R - 2\lambda)\cr
&= 4\pi\gamma\chi - 2 \lambda \int \sqrt{{-\bar g}}e^{2\sig},\cr}
\equn\put\stwo
$$
where $\chi$ is the Euler number and in the second line of {\stwo} we
used the conformal gauge {\confdef}. To avoid possible confusion due to
the triviality of pure gravity in two-dimensions, we consider also the
presence of $d$ matter fields. The classical theory appears then to
have $d-2$ dynamical degrees of freedom, in view of the equations of
motion: $T^{ab}_{\rm matter}=\lambda g^{ab}$; two of these equations
act as constraints while the third one, corresponding to the trace,
determines $\sig$ in terms of the matter fields. At the quantum level
the dynamics of the theory changes discontinuously because of the
trace anomaly. In fact the trace of the classical equations of motion,
which corresponds precisely to the $\sig$-variation, cannot be preserved
similtaneously with general coordinate invariance. For conformally
invariant matter, the trace anomaly takes the form:
$$
\eqalign{T &= {c\over 24 \pi} R \cr
&= {c\over 24 \pi} e^{-2 \sig} (\bR - 2\sqb\sig), \qquad D=2\cr}
\equn\put\anomtwo
$$
in the decomposition {\confdef}, where $T\equiv T_a^a$. The central
charge $c$ is given by: $$
c=d-26+1, \qquad d=N_S + N_F,
\equn\put\ctwo
$$
where -26 and +1 stand for the contribution of the reparametrization
ghosts and $\sig$-field itself, and the contribution to the matter
central charge $d$ of $N_S$ free scalars and $N_F$ free (Dirac) fermions
is also presented. There are two ways to define a consistent quantum
theory:
\hfill \break
(i) Cancel the conformal anomaly when $d=26$, in which
case $\sig$ decouples and one is left with critical string theory. The
classical counting of $d-2$ degrees of freedom remains valid.
\hfill \break
(ii) Modify the classical action {\stwo} by adding the
anomaly-induced Polyakov term [\cPoly]:
$$
\eqalign{S^{(2)}_{\rm anom} &= -{c\over 96 \pi} \int \sqrt{-g}\
R{1\over \sq}R \cr &= {c\over 24 \pi} \int \sqrt{-\bar g} (-\sig \sqb
\sig +\bR \sig ), \cr}
\equn\put\satwo
$$
such that
$$
T = {1 \over \sqrt{- g}}{\delta\over\delta\sig}S^{(2)}_{\rm anom},
\equn\put\sadef
$$
and in the second line of {\satwo} a $\sig$-independent term has been
dropped. In the resulting Liouville action, the conformal factor has now
two derivatives and acquires its own dynamics. One is left with
non-critical string theory describing $d-1$ physical propagating modes.

Let us now consider the same line of reasoning in four dimensions. The
classical action is:
$$
\eqalign{S^{(4)}_{cl} &= {1\over 2\kappa}\int\sqrt{-g} (R-2\Lambda )\cr
&={1 \over 2 \kappa} \int \sqrt{-{\bar g}}\{ e^{2 \sig}
(\bR + 6 (\nabar \sig)^2 ) - 2\Lambda e^{4\sig}\} , \cr}
\equn\put\sfour
$$
where $\kappa = 8 \pi G_N $. Although $\sig$ appears with two derivatives
in {\sfour}, it is still non-propagating due to the reparametrization
constraints. On the other hand, the metric ${\bar g}_{ab}$ in the
decomposition {\confdef} cannot be fixed by a gauge choice as in two
dimensions, since it contains the two physical transverse, trace-free,
spin-2 excitations. At the quantum level, the general form of the
trace anomaly for conformally invariant matter in a four dimensional
curved space-time is a linear combination of the square of the Weyl
tensor $C^2$, the Gauss-Bonnet combination $G$, and $\sqR$ [\cDuf]:
$$
\eqalign{T &= bC^2 +  b'(G-{2 \over 3}\sqR ) + {\zeta\over 3}\sqR ;
\qquad D=4 \cr
C^2 &= \Rie - 2 \Ric + {1 \over 3} \R ,\cr
G &= \Rie - 4 \Ric + \R .\cr}
\equn\put\anomf
$$
For free fields, the anomaly coefficients $b$ and $b'$ are given by:
$$
\eqalign{b &= {1 \over 120 (4 \pi)^2} (N_S + 6N_F + 12N_V -8),\cr
b'&= -{1 \over 360 (4 \pi)^2} (N_S + 11N_F + 62N_V -28),\cr}
\equn\put\bcoef
$$
with $N_V$ the number of vectors and the last numbers denote the
$\sig$-contribution [\cAMM]. Unlike $b$ and $b'$, the coefficient of the
$\sq R$ term in the anomaly is altered by the addition of a {\it local}
$R^2$ term in the action, whose conformal variation is exactly $\sq R$,
so that it must be treated as an additional renormalized coupling, and
the $\zeta$ coefficient in {\anomf} is left undetermined. In addition to
the $C^2$, $G$ and $\sq R$ terms in the general form of the  trace
anomaly for $D = 4$, an $R^2$ term is allowed by naive dimensional
analysis. However, such a term is forbidden by the Wess-Zumino
consistency condition [\cwz], which in the present context is simply the
statement that the variational relation {\sadef} is integrable, {\it
i.e.} that there exists an effective action functional (local or
non-local) $S_{\rm anom}$, whose $\sig$ variation is the anomaly.

In the parametrization {\confdef}, the quantity $\sqrt{-g} C^2$ is
independent of $\sig$, while the combination $\sqrt{-g}(G-{2 \over 3}
\sqR)$ becomes only linear in $\sig$:
$$
\sqrt{-g}(G -{2 \over 3}\sqR) = \sqrt{-{\bar g}}(4{\overline \Delta}\sig
+ {\overline G} - {2\over 3} \sqb \bR ),
\equn\put\gbcomb
$$
where $\Delta$ is the Weyl covariant fourth order operator acting on
scalars:
$$
\Delta = \sq ^2 + 2R^{ab} \nabla_a
\nabla_b - {2 \over 3} R {\sq}  + {1 \over 3} (\nabla^a R) \nabla_a .
\equn\put\deltaf
$$
Thus, as in two dimensions, the $\sig$ dependence of the non-local
anomaly-induced action becomes local in the conformal parameterization
{\confdef}:
$$
\eqalign{S^{(4)}_{\rm anom} &= - \int\sqrt{-g}\{ {1 \over 8b'}
[ b C^2 +b'(G-{2\over 3}\sqR)] {1 \over\Delta}  [b C^2 + b'(G-{2\over
3}\sqR )] - {\zeta \over 36} R^2 \}  \cr
&= \int\sqrt{-{\bar g}}\{ 2b'\sig \overline\Delta \sig +
[b{\overline C}^2 + b'(\overline G - {2\over 3} \sqb \bR )]\sig
-{\zeta \over 36}[\bR - 6\sqb\sig - 6(\nabar \sig)^2 ]^2 \},\cr}
\equn\put\safour
$$
where in the second line of {\safour} a $\sig$-independent term has been
dropped.

This action plays a role similar to the Wess-Zumino action of low energy
pion physics, as realized in the Skyrme model, for example. That is, it
can be interpreted as an effective action at low energies, which
describes all modifications to the Ward-identites due to the presence of
the quantum trace anomalies. In particular, it guarantees that
the total trace anomaly vanishes at any fixed point of the matter
theory, provided that the coefficients $b$, $b'$ and $\zeta$ are chosen
appropriately. Using in fact the integrability condition {\sadef}, one
has the identity:
$$
S_{\rm matter}[g_{ab}= e^{2 \sig} \bar g_{ab}] = S_{\rm matter}[ \bar
g_{ab}]  + S_{\rm anom} [\bar g_{ab}; \sig].
\equn\put\wz
$$
Both sides of the above equation are trivially invariant under the
transformation
$$
\eqalign{\bar g_{ab} &\rightarrow e^{2 \omega}\bar g_{ab}\cr
\sig &\rightarrow \sig - \omega, \cr}
\equn\put\weyl
$$
which leaves the total metric unchanged. The infinitesimal variation of
{\wz} with respect to $\omega$ then yields:
$$
T_{\rm matter} + T_{\sig} = {\delta S_{\rm anom}\over \delta\sig},
\equn\put\tzero
$$
implying that the total trace vanishes when one uses the $\sig$-equation
of motion.

In the induced action $\safour$ $\sig$ appears with four derivatives, as
opposed to two in the classical theory {\sfour}, indicating the
presence of an extra mode. This raises the question whether any
consistent theory of gravity requires the existence of an additional
scalar field with the properties of $\sig$. The answer is not obvious
because an essential difference between four and two dimensions is that
the classical theory is non-renormalizable, implying that not only the
trace, but all the components of the classical equations of motion are
in principle inconsistent at the quantum level. Note however that in
string theory, which is the only known example of quantum gravity,
there is such a massless scalar mode, the dilaton, which at the lowest
order couples to the Gauss-Bonnet integrand as $\sig$ and can be used to
cancel the trace anomaly. For our purposes, we are interested in the
effective theory at very large distances, where the ultraviolet
structure of quantum gravity is not relevant. On the other hand,
anomalies correspond to non-local terms which can not be removed by
local counter terms and, therefore, they are important at all scales and
have to be taken into account.
In particular, the quantum trace anomaly {\anomf} modifies precisely the
trace of the classical equations of motion of Einstein's theory {\sfour},
which, for conformally invariant matter, fixe the value of the curvature
scalar $R=\Lambda$. In the induced $\sig$-action, $R$ appears with
derivatives and the conformal factor aquires non-trivial dynamics which
in principle could distabilize the classical vacuum.

When $\bar g_{ab}$ is conformally flat, the effective $\sig$-action may
be derived by a completely different method, based on symmetry
argument. In fact, conformal flatness implies that the fiducial metric
has a set of $D + 1$ conformal Killing vectors, denoted generically by
$\xi_a$, satisfying:
$$
({\overline L}\xi)_{ab} \equiv \nabar _a \xi_b + \nabar _b
\xi_a - {2 \over D}\bar g_{ab} \nabar\cdot \xi = 0.
\equn\put\ckv
$$
As a result, the effective $\sig$-theory has a residual {\it
global} conformal ivariance, which is a remnant of the coordinate
invariance of the full theory. It arises as a combination of the
coordinate transformation $x^a\rightarrow x^a +\xi^a$ and the rescaling
{\weyl} with $\omega =-{2\over D} \nabar\cdot\xi$, which leaves the
metric $\bar g_{ab}$ invariant. It is now easy to show that
{\stwo}+{\satwo} or {\sfour}+{\safour} is the most general local action
for $\sig$ in two or four dimensions containing up to two or four
derivatives, respectively, wich is invariant under the global conformal
transformations: $\delta\sig = \xi\cdot\nabar\sig +{1\over D}
\nabar\cdot\xi$. Under this transformation, although $\sig$ transforms
inhomogeneously, the conformal factor $e^{\sig}$ has scaling dimension
equal to unity (as does $\partial\sig$).
\vskip 1.0cm
\centerline{\bf Quantization of $\sig$-theory}
\vskip 0.5cm

When the conformal factor is quantized, the vanishing of the total
trace at the quantum level requires to find a non-trivial critical
behaviour at the infrared of the effective $\sig$-theory. Remarkably,
the effective action {\safour}+{\sfour} is ultraviolet renormalizable
and the various beta functions may be studied in ordinary flat space
perturbation theory. The $\zeta$ coupling is infinitely  renormalized,
and contributes an $R^2$ term to the anomaly, proportional to its
$\beta$-function, $\beta_{\zeta}$. Such a term is forbiden by the
Wess-Zumino consistency condition, implying that this $\beta$-function
must vanish. By simple power counting, it follows that
$\zeta$-renormalization is not affected by the interactions of {\sfour},
and a straight-forward one-loop calculation shows that $\zeta =0$ is an
infrared stable perturbative fixed point [\cAM]. This condition fixes
the arbitrariness present in $\zeta$, the coefficient of the local $R^2$
term, which is usually present in higher-derivative theories of gravity.
The induced action {\safour} is then given entirely by the non-local
term, in analogy with the Polyakov action {\satwo} in two dimensions.
In conformal coordinates {\confdef}, it becomes:
$$
S^{(4)}_{\rm anom} = \int\sqrt{-\bar g}\{ -{Q^2 \over (4 \pi)^2}[\sig
\overline\Delta \sig +{1\over 2}(\overline G - {2\over 3}\sqb \bR )]\sig
+ b{\overline C}^2\sig \},
\equn\put\sanl
$$
where $Q^2 =-32\pi^2 b'$. Recall that $b'$ was the coefficient of the
Gauss-Bonnet term in the trace anomaly {\anomf}, which is negative
definite for all known matter fields {\bcoef} corresponding to a bounded
Euclidean action. This is in contrast to the two dimensional case
{\ctwo}, where the matter contributes to a negative kinetic term for
the Liouville theory {\satwo}. This quantity has been proposed as the
four dimensional analog of two dimensional central charge, $c$, for
which the Zamolodchikov theorem applies [\cZam], namely it decreases
monotonically under the renormalization group flow from an ultraviolet
to an infrared fixed point.
At $\zeta =0$, the contribution to the anomaly coefficients $b$ and $b'$
of the $\sig$-field itself is coming from the quartic operator $\Delta$
{\deltaf} and was computed in ref.[\cAMM]; the results are quoted in
{\bcoef}.

The effective action {\sanl} involves four derivatives of $\sig$ and
raises the problem of unitarity, known to plague local higher derivative
theories. However, recall that the Einstein action in its covariant
form {\sfour} appears to contain a negative metric scalar degree of
freedom, but that this state is removed by the reparameterization
constraints. In the quartic action {\sanl} there remains in principle
one additional scalar dilaton mode not cancelled by the constraints.
It is instructive to consider background metrics parametrizing a general
Einstein space, ${\overline R}_{ab}={{\overline R}\over 4}{\bar
g}_{ab}$, in which case the quartic operator $\overline\Delta$
factorizes into the product of two second order operators: $\Delta
=\sq(\sq -{R\over 6})$. This shows that the quartic action may be
regarded as containing a conformally coupled scalar mode from $\sq
-{R\over 6}$ which has a negative kinetic term, and a minimally coupled
scalar from $\sq$ with a positive kinetic term. The first state is
identical to the one of Einstein gravity {\sfour} and must be cancelled
by the ghosts if the theory is to be unitary, in analogy with the
Liouville mode in two dimensions when $d>25$. The operator $\sq
-{R\over 6}$ is the precise analog of the scalar Laplacian in two
dimensions which contributes to the central charge like one additional
scalar degree of freedom {\ctwo}. The minimally coupled scalar mode with
positive kinetic term has no analog in two dimensions or the Einstein
theory. It makes negative the full contribution of $\Delta$ to the
central charge {\bcoef} and is responsible for the unusual behavior of
this theory in the infrared, as will be discussed below [\cAM].

We now turn to the problem of the ghost, as well as the transverse
spin-2 contribution to the effective theory of $\sig$. As already
mentioned in the introduction, these graviton modes will be treated as
the remaining massless matter fields, which amounts to evaluate their
own contribution to the anomaly coefficients. This approximation is
justified if the quantum fluctuations of transverse modes can be
neglected at such large distances we consider. Note that this is not the
case in the quadratic approximation of the Einstein theory around de
Sitter space, where it was found that the transverse, traceless part of
the graviton propagator grows logarithmically at large distances. Such a
behaviour results from taking the inverse of the corresponding kinetic
operator $-\sq +{R\over 6}$. However, when the $\sig$-equation of motion
which it is associated to the trace anomaly is not used, the kinetic
operator becomes $-\sq +{2\over 3}R-2\Lambda$ whose inverse has not
singular infrared behaviour for generic values of $\Lambda$.

Although no particular gauge choice is preferred over any other, a
convenient and geometric way to compute the ghost and spin-2 contribution
to the trace anomaly is by factorizing from a general metric the
diffeomorphisms and Weyl rescalings, in analogy with two dimensions
[\cPoly,\cmes]:
$$
g_{ab}(x) = {\partial X^{\mu}\over \partial x^a}
{\partial X^{\nu}\over \partial x^b} e^{2\sigma (X)}
g_{\mu\nu}^{\perp}(X),
\equn\put\dec
$$
where $g_{\mu\nu}^{\perp}$ denotes the transverse trace-free part. The
covariant measure on the function space of metrics is defined by
means of the DeWitt supermetric on this space. The change of variables
from $g_{ab}$ to $\sig$ and $g_{\mu\nu}^{\perp}$ after dividing by
the volume of diffeomorphisms, results in a Jacobian factor in the
measure:
$$
J = {\det}^{'{1 \over 2}} (L^{\dag} L),
\equn\put\jac
$$
where $L^{\dag}$ is the Hermitian adjoint of $L$ as defined in {\ckv}.
The prime in {\jac} indicates that the zero modes of $L$ must be excluded
from $J$ and treated separately.

$L^{\dag}L$ is the product of two operators each of which transform
covariantly under a local conformal transformation:
$$
L = e^{2 \sig} {\overline L} e^{-2 \sig}\qquad , \qquad
L^{\dag} = e^{-D\sig} {\overline L}^{\dag} e^{(D-2)\sig},
\equn\put\Ltrans
$$
under the substitution {\confdef}. Using the heat kernel definition for
the determinant and these transformation properties we find:
$$\eqalign{
-{1 \over 2}\delta \ln {\det}'  (L^{\dag}L) &= {1\over 2}{\trp}\delta
\int_{\eps}^{\infty}{ds \over s} e^{-s L^{\dag}L}\cr &=
-{D+2 \over 2}{\trp} \delta\sig e^{-\eps  L^{\dag}L}  +
{D \over 2}{\trp} \delta\sig e^{-\eps LL^{\dag}},\cr }
\equn\put\var
$$
where the cyclic property of the trace has been used repeatedly, and
the lower limit of the proper time heat kernel has been regulated
by $\eps$, to be taken to zero in the end. Because of the explicit
appearance of ${\trp}$ over the subspace of nonzero modes of $L$, the
upper limit of the evaluation of the integral in $s$ vanishes and only
the lower limit survives in {\var}. Here, an essential difference from
the two dimensional case manifests itself in the appearance of the {\it
tensor} operator $LL^{\dag}$ whose kernel is {\it infinite} dimensional,
being spanned by all transverse, traceless graviton mode fluctuations.
Unlike for $D= 2$, where the zero modes of $LL^{\dag}$ are countable by
the Riemann-Roch theorem, and their conformal variations may be added
explicitly to {\var}, in $D=4$ these modes cannot be ``counted" without
some action over the transverse, traceless degrees of freedom.
Equivalently, if we exclude these modes by restriction to the non-zero
mode space of $LL^{\dag}$, then the conformal variation of the ghost
operator $L^{\dag}L$ in {\var} is necessarily non-local, and violates
the Wess-Zumino consistency by itself. The ghost operator cannot yield a
coordinate invariant effective action unless it is combined with the
action for the physical graviton modes.

Considering the Einstein action for the spin-2 degrees of freedom,
we run into several difficulties. On the one-hand, this theory
is not classically conformally invariant. On the other hand, loop
calculations in quantum gravity, as in non-Abelian gauge theory, make
sense only if the the  background field equations are satisfied.
However, the field equations of the Einstein theory obscure the WZ
condition, since they imply  that any $R^2$ term in the anomaly is
indistinguishable from $\Ric$, on shell. Another example is the
Weyl-squared action, which has the advantage of being classically
conformally invariant, so that the WZ condition may be checked
explicitly. Moreover, the restriction to the one-loop contribution of
this action is equivalent to imposing a self-duality constraint on the
graviton degrees of freedom [\cselfd]. Of course, use of this higher
derivative action for the graviton modes leads to perturbative
non-unitarity, about which we have nothing new to add. If we simply
ignore these difficulties, and perform the calculation in the weak-field
limit, one finds that actually the numerical results for these two
actions are not significantly different [\cAMM]:
$$\eqalign{
b_{\rm grav} &= {1\over (4 \pi)^2}\ {\ 611\over 120}, \qquad
b'_{\rm grav} = -{1\over (4 \pi)^2}\ {1411\over 360} \qquad
({\rm Einstein}) \cr
b_{\rm grav} &= {1\over (4 \pi)^2}\ {199\over 30},\ \qquad
b'_{\rm grav} = -{1\over (4 \pi)^2}\ {\ 87\over 20}\ \ \qquad
({\rm Weyl}) \cr}
\equn\put\qgrav
$$
This gives a positive contribution to $Q^2$ of 7.9 or 8.7 for the
Enstein or Weyl theory, respectively. In either case, it is noteworthy
that the total $b$ and $-b'$ coefficients are positive, and dominated by
the ghost + graviton contributions, which add with the same sign as the
matter contributions. This is different from the two dimensional result
that the matter and ghosts contribute to the central charge with opposite
sign.
\vskip 1.0cm
\centerline{\bf Anomalous scaling behaviour}
\vskip 0.5cm

Once the classical action {\sfour} is added to the induced action
{\sanl}, one obtains a four-dimensional analog of the Liouville theory
{\satwo}+{\stwo}. The exponential interactions of {\stwo} for $D=2$, or
{\sfour} for $D=4$, are classically conformally invariant with the
conformal factor $e^{\sig}$ having scaling dimension equal to unity.
When $\sig$ is quantized, there is anomalous scaling behaviour which can
be determined by the requirement of vanishing $\beta$-functions for the
couplings of the exponential interactions. In fact, the Liouville theory
is superrenormalizable and, thus, the exact scaling dimensions can be
computed in ordinary perturbation theory by analyzing only a finite set
of divergent diagrams. A convient way to do the calculation is to assume
that $e^{\sig}$ acquires a scaling dimension $\alpha$ and to define
$\sig=\alpha{\hat\sig}$, so that the rescaled field $e^{\hat\sig}$ has
weight one.

To illustrate the idea let us consider first the two-dimensional case
[\cDDK]:
$$
{\cal L}_{\rm eff}^{(2)} = -{Q^2\over 4\pi}[(\nabar\sig )^2
+\bR\sig ] -2\lambda e^{2\sig},
\equn\put\lefft
$$
where $Q^2 ={1\over 12}(25-d)$. Ordinary power counting implies that
primitive divergences arise only from tadpoles, in which the coupling
$\lambda$ appears exactly once. Substituting $\sig=\alpha{\hat\sig}$,
and varrying the coupling $\lambda$ with respect to some mass scale,
one finds that its $\beta$-function is:
$$
\beta_{\lambda} = (2-2\alpha + {\alpha^2\over Q^2})\lambda ,
\equn\put\bltwo
$$
where the first two terms in the r.h.s. of {\bltwo} represent the
classical contribution, while the third term is the quantum contribution
from the one-loop tadpole graph. The vanishing of this beta function for
$\lambda \neq 0$ yields a quadratic relation for the anomalous scaling
of $e^{\sig}$:
$$
\alpha = {1-\sqrt{1 - {2\over Q^2}} \over {2\over Q^2}},
\qquad\qquad D=2
\equn\put\alpt
$$
where we have chosen the negative branch of the square root, so that the
classical scaling $\alpha =1$ is obtained in the limit $Q^2\rightarrow
\infty$. The critical exponent {\alpt} is real only for $Q^2\ge 2$
corresponding to $d\le 1$.

Going back to four dimensions, one has [\cAM]:
$$
{\cal L}_{\rm eff}^{(4)} =
-{Q^2\over (4\pi )^2}[\sig\sqb (\sqb -{\bR\over 6}) \sig
+{\bR^2\over 12}\sig ] + \gamma e^{2 \sig}[(\nabar \sig)^2 +{\bR\over
6}]  -\lambda e^{4\sig},
\equn\put\lefff
$$
where $\gamma ={3\over\kappa}$, $\lambda ={\Lambda\over\kappa}$, and
for simplicity we consider as backgrounds Einstein spaces with
vanishing Weyl-squared; these include maximally symmetric spaces, as de
Sitter space-time. It is easy to show that the four-dimensional
Liouville-like theory {\lefff} is superrenormalizable because of the
quartic propagator. In this case, primitive divergences arise not only
from tadpoles involving the $\gamma$ or $\lambda$ couplings, but also
from graphs containing two $\gamma$-vertices [\cAM]. However, the
renormalization of $\gamma$-coupling is multiplicative and arises only
from tadpoles, leading to the following $\beta$-function:
$$
\beta_{\gamma} = (2-2\alpha + 2{\alpha^2\over Q^2})\gamma ,
\equn\put\bgfour
$$
which is analogous to {\bltwo} in two dimensions. Its vanishing for
$\gamma \neq 0$ yields the anomalous scaling dimension of the
conformal factor $e^{\sig}$:
$$
\alpha = {1-\sqrt{1 - {4\over Q^2}} \over {2\over Q^2}},
\qquad\qquad D=4.
\equn\put\alpf
$$
The value $Q^2 = Q_{cr}^2 = 4$ corresponds to $d=1$ in two dimensions,
where the theory could exhibit a phase transition or qualitatively new
phenomena. However, it seems from {\bcoef} and {\qgrav} that the
physically relevant case in four dimensions is always $Q^2 > 4$,
corresponding to $d < 1$ in two dimensions. Finally, $\lambda$ coupling
is not multiplicatively renormalized, but it mixes with $\gamma$ from
non-tadpole diagrams. Its $\beta$-function reads:
$$
\beta_{\lambda} = (4 - 4\alpha + {8\alpha^2 \over Q^2}) \lambda
- {8 \pi^2 \alpha^2 \over Q^4} \gamma^2
( 1 + {4 \alpha^2 \over Q^2} + {6 \alpha^4 \over Q^4}).
\equn\put\blfour
$$
Since $\alpha$ has already been determined by {\alpf}, setting
$\beta_{\lambda} = 0$ gives a non-trivial relation for the cosmological
constant $\lambda$ in Planck ``units" $\gamma$, which is a function of
the ``central charge" $Q^2$.

There is an equivalent way of deriving the scaling relation {\alpf} by
requiring that the operator, $\sqrt{-g} R$ has conformal weight equal to
four. For this, one must use that the dimension of the operator
$e^{p\sig}$ is
$$
[e^{p\sig}]= p - {p^2 \over 2 Q^2},
\equn\put\dim
$$
where $p$ is its classical value and the second term represents the
quantum contribution. Once this condition has been imposed, note that
one can no longer insist that the cosmological term $e^{4\alpha\sig}$
have the same conformal weight.  Instead, there is a non-trivial mixing
between the $\lambda$ and $\gamma$ couplings, so that invariance can be
enforced only by the relation $\beta_{\lambda} = 0$ in {\blfour}.

The above calculations, we performed using the short-distance behaviour
of the theory in flat space, can also be carried out in the infrared
around de Sitter space yielding the same $\beta$-functions. In fact, the
propagator corresponding to the quartic term, $\sqb (\sqb - {\bR \over
6})$ is dominated by the (non unitary) $\sqb^2$ term at short distances,
and by the (unitary) $-{\bR \over 6}\sqb$ term at large distances.
However, in de Sitter space [\cAntMot],
$$
\sqb_{xx'}^{-1} = -{1 \over 8 \pi^2} [{2 \over s^2(x,x')} - H^2 \ln
({H^2 s^2 (x,x') \over 4})], \qquad \bR = 12 H^2,
\equn\put\boxds
$$
where $s(x,x')$ is the geodesic distance between the two points. Hence,
the quartic propagator has the {\it same} logarithmic behavior in both
limits. This is not surprising, as it is known from critical phenomena
that there is a close interplay between ultraviolet and infrared
behavior in systems with conformal symmetry.

We now observe that if the expectation value of the Ricci scalar is
different from zero, then the global conformal symmetry must be
spontaneously broken. In fact in the semi-classical limit, when the
anomaly induced fluctuations are suppressed by $ 1 \over Q^2$ for large
$Q^2$, the dimension of the conformal factor $\alpha =1$ and the weight
of $R$ is zero, {\it i.e.} it transforms like a scalar under global
conformal transformations. However, this is not the case at the
non-trivial fixed point {\alpf}, where $\alpha\ne 1$ and the conformal
weight of $R$ is not zero:
$$
[R] = -2 \alpha - {2 \alpha^2 \over Q^2} + 2 = 4 (1 - \alpha),
\equn\put\dimr
$$
where we used {\dim}. This implies that $\langle R\rangle$ becomes an
order parameter for the spontaneous breaking of global conformal
symmetry, in sharp contrast to the classical situation in which $\langle
R\rangle$ can take on any value consistent with the symmetry. As a
result, the cosmological ``constant" problem reduces to the question of
whether this symmetry remains spontaneously broken, or is restored in
the quantum theory.

We may compare this case to that of spontaneous breaking of a continuous
symmetry in two dimensions [\cBer]. Consider a complex scalar field $\phi
(x)$ with a tree-level potential giving rise to symmetry breaking. When
the field is quantized, the corresponding massless Goldstone boson has a
propagator which grows logarithmically at large distances. This infrared
divergence implies instability of the spontaneously broken vacuum due to
quantum fluctuations. Because of this instability of the ordered state,
the system becomes disordered and the $U(1)$ symmetry is restored at the
quantum level. Locally, there are regions of broken symmetry in which
the classical description remains valid. However, as we consider
regions of larger and larger size, the classical description breaks
down and the average expectation value vanishes. The quantitative
description of this phenemenon is
given by the power law fall-off of the correlation function
$\langle\phi (x)\phi^{\dag}(0)\rangle$. Introducing the nonlinear
polar field decomposition $\phi = \rho e^{i\theta}$ and neglecting
the fluctuations of the massive $\rho$ field in the infrared, one finds
that the angular field $\theta$ may be treated as a free field
with the propagator:
$$
\langle\theta (x) \theta (0)\rangle = -{1 \over 4\pi\rho^2} \ln (\mu^2
x^2).
\equn\put\phase
$$
In this infrared scaling limit, the correlation function for $\phi$ has a
power law behavior:
$$
\eqalign{\langle\phi (x) \phi^{\dag}(0)\rangle &\sim \rho^2
\langle e^{i \theta (x)} e^{-i \theta (0)}\rangle \cr
&\sim \rho^2 e^{\langle\theta (x) \theta (0)\rangle} \
\sim \rho^2 \ |x|^{-{1 \over 2\pi\rho^2}}.\cr}
\equn\put\phiphi
$$

Consider now by analogy the correlation function of Ricci scalars,
$\langle R(x) R(x') \rangle$ at two different points. Using {\confdef}
with $\sig$ replaced by $\alpha \sig$, and the $\sig$-propagator from
{\lefff} and {\boxds}, we find at large distances $|s(x,x')| \rightarrow
\infty$:
$$
\eqalign{\langle R(x) R(x') \rangle
&\sim H^4\langle e^{-2\alpha\sig (x)}e^{-2\alpha\sig (x')}\rangle \cr
&\sim H^4 e^{4\alpha^2 \langle\sig (x)\sig (x')\rangle} \
\sim H^4 \ |H s(x,x')|^{-4{\alpha^2 \over Q^2}}\ ,\cr}
\equn\put\cosm
$$
where only the dominant infrared behavior has been retained.
The result {\cosm} states that the {\it effective} cosmological
``constant" goes to zero with a definite power law behavior for large
distances. In other words, there is screening in the infrared of the
effective value of vacuum energy at larger and larger scales. In
contrast with the 2D case {\phiphi}, the value of the power is {\it
universal}, depending only on $Q^2$ which counts the effective number of
massless degrees of freedom. In particular, it depends neither on the
classical value of the background curvature $\bR$, nor on the Planck
scale. This is essential for a scale invariant, phenemenologically
acceptable solution of the cosmological constant problem.

To summarize, we believe that the effective theory of the conformal
factor presented here provides a useful framework for studying the
infrared behavior of gravity in four dimensions and addressing the
cosmological constant problem. The anomalous scaling of the conformal
factor may be the key to understanding why $\langle R \rangle = 0$ in
the observed universe. However, many unanswered questions and open
problems remain. An explicit proof of the unitarity of the
$\sig$-theory when one applies the diffeomorphism constraints, an
explicit verification of the approximation of neglecting the contribution
of transverse spin-2 modes in the infrared, a better understanding of
the correlation functions and their scaling behavior at large
distances, and observational implications to cosmology, large scale
structure and the missing matter puzzles.
\baselineskip = 15pt
\parindent=-10 pt

\vskip 1.0cm
\centerline{\bf References}
\vskip 0.5cm
\parskip=-3 pt

\item{\cgrav}. L.H. Ford, {\it Phys. Rev.} {\bf D31} (1985) 710;
I. Antoniadis, J. Iliopoulos, T.N. Tomaras, {\it Phys. Rev. Lett.}
{\bf 56} (1986) 1319; B. Allen and M. Turyn, {\it Nucl. Phys.}
{\bf B292} (1987) 813; E.G. Floratos, J. Iliopoulos, and T.N. Tomaras,
{\it Phys. Lett.} {\bf B197} (1987) 373. \hfill\break

\item{\cAntMot}. I. Antoniadis and E. Mottola, {\it Jour. Math. Phys.}
{\bf 32} (1991) 1037. \hfill\break

\item{\cPoly}. A. M. Polyakov, {\it Phys. Lett.} {\bf B103} (1981)
 207. \hfill\break

\item{\cAM}. I. Antoniadis and E. Mottola, {\it Phys. Rev.} {\bf D45}
(1992) 2013. \hfill\break

\item{\cAMM}. I. Antoniadis, P.O. Mazur and E. Mottola, Ecole
Polytechnique preprint CPTH-A173.0492, to appear in {\it Nucl. Phys.}
{\bf B}. \hfill\break

\item{\cselfd}. C. Schmidhuber, CalTech report CALT-68-1745
(1992). \hfill\break

\item{\cDuf}. M.J. Duff, {\it Nucl. Phys.} {\bf B125} (1977) 334;
N. D. Birrell and P. C. W. Davies, {\it
``Quantum Fields in Curved Space,"} (Cambridge Univ. Press,
Cambridge, 1982) and references therein. \hfill\break

\item{\cwz}. J. Wess and B. Zumino, {\it Phys. Lett.} {\bf B37} (1971)
95; L. Bonora, P. Cotta-Rasmusino, and C. Reina, {\it Phys. Lett.}
{\bf B126} (1983) 305. \hfill\break

\item{\cZam}. A.B. Zamolodchikov, {\it JETP Lett.} {\bf 43} (1986) 730;
{\it Sov. J. Nucl. Phys.} {\bf 46} (1987) 1090 [{\it Yad. Fiz.} {\bf 46}
(1987) 1819]; J. L. Cardy, {\it Phys. Lett.} {\bf B215} (1988) 749;
A. Cappelli, D. Friedan and J.I. Latorre, {\it Nucl. Phys.} {\bf B352}
(1991) 616. \hfill\break

\item{\cmes}. P.O. Mazur and E. Mottola, {\it Nucl. Phys.} {\bf B341}
(1990)
187; S.K. Blau, Z. Bern, and E. Mottola, {\it Phys. Rev.} {\bf D43}
(1991) 1212. \hfill\break

\item{\cDDK}. F. David, {\it Mod. Phys. Lett.} {\bf A3} (1988) 1651;
J. Distler and H. Kawai, {\it Nucl. Phys.} {\bf B321} (1989) 509.
\hfill\break

\item{\cBer}. N.D. Mermin and Wagner, {\it Phys. Rev. Lett.} {\bf 17}
(1966) 1133; V.I. Berezinskii, {\it Zh. Eksp. Teor. Fiz.} {\bf 59}
(1970) 907 \ [{\it Sov. Phys. JETP} {\bf 32} (1971) 493];
S. Coleman, {\it Comm. Math. Phys.} {\bf 31} (1973) 259;
S.K. Ma and R. Rajaraman, {\it Phys. Rev.} {\bf 11} (1975) 1701.
\hfill\break

\end